\documentclass[%
 reprint,
superscriptaddress,
nofootinbib,
 amsmath,amssymb,
 aps,
]{revtex4-2}

\usepackage{graphicx}
\usepackage{dcolumn}
\usepackage{bm}
\usepackage{hyperref}
\usepackage{natbib}
\usepackage{cleveref}
\usepackage{mathbbol}
\usepackage{mathtools}
\usepackage{xcolor}
\usepackage{dsfont}
\usepackage{comment}
\usepackage[normalem]{ulem}
\usepackage[pagewise,mathlines]{lineno}

\begin{document}


\title{Variational approach to nonholonomic and inequality-constrained mechanics}

\author{A. Rothkopf}
\email{akrothkopf@korea.ac.kr}
\affiliation{%
Department of Mathematics and Physics, University of Stavanger, Kristine Bonnevies vei 22, 4021 Stavanger, Norway
}%
\affiliation{%
Department of Physics, Korea University, Seoul 02841, Republic of Korea
}%
\author{W. A. Horowitz}
\email{wa.horowitz@uct.ac.za}
\affiliation{Department of Physics, University of Cape Town, Private Bag X3, Rondebosch 7701, South Africa}%
\affiliation{Department of Physics, New Mexico State University, Las Cruces, New Mexico, 88003, USA}
\affiliation{Theoretical Sciences Visiting Program, Okinawa Institute of Science and Technology Graduate University, Onna, 904-0495, Japan}

\begin{abstract}
Variational principles play a central role in classical mechanics, providing compact formulations of dynamics and direct access to conserved quantities. While holonomic systems admit well-known action formulations, non-holonomic systems -- subject to non-integrable velocity constraints or position inequality constraints -- have long resisted a general extremized action treatment. In this work, we construct an explicit and general action for non-holonomic motion, motivated by the classical limit of the quantum Schwinger-Keldysh action formalism, rediscovered by Galley. Our formulation recovers the correct dynamics of the Lagrange-d'Alembert equations via extremization of a scalar action. We validate the approach on canonical examples using direct numerical optimization of the novel action, bypassing equations of motion. Our framework extends the reach of variational mechanics and offers new analytical and computational tools for constrained systems.
\end{abstract}

\maketitle


\section{Introduction}

The 17th century saw the emergence of the foundations of modern mechanics, the study of motion and interaction among physical objects. In fact two complementary approaches were put forward by Newton and Leibnitz concurrently \cite{lanczos2012variational}. While Newton focussed on the study of forces and the resulting changes in momentum, Leibnitz studied what we would now call work and the resulting changes in energy. Leibnitz' formulation of mechanics in terms of energy laid fertile ground for Lagrange \cite{lagrange1788mechanique} and eventually Hamilton \cite{hamilton1834general}  to uncover a key property of a vast class of mechanical systems: their motion is optimal in that the path taken by the physical objects extremizes the difference between kinetic $T$ and potential energy $V$ along the way. Formulated in terms of the Lagrange function $L=T-V$ one can construct the so-called action functional $S=\int dt \, L$. For any trajectory the system may take, $S$ produces a number (a score), which for the physical trajectory becomes a saddle point\footnote{Only for small enough times and before the occurrence of turning points, does the classical trajectory minimze the action.}. This action principle of Hamilton amounts to a highly efficient representation of all information needed to study system motion. Only through the action principle was it possible for Emmy Noether to reveal \cite{noether1918invariante} that all conservation laws follow from symmetries, laying the ground work for the symmetry revolution in 20th century physics \cite{gross96}.

While offering compelling conceptual and technical benefits, a key limitation of Hamilton's action principle is that it only applies to systems in which conservative forces act. I.e.\ work done by the forces in the system may only depend on initial and final point of motion. Even more importantly for practical application, Hamilton's principle only allows us to include constraints to system motion that at most depend on the position of the moving object, such as when an ideal pendulum is constrained to move on a circular path. Constraints related to elementary types of motion, such as those arising in the motion of wheels, on the other hand depend on the velocity of the object in motion in a non-linear fashion and foil Hamiton's principle.

In fact the vast majority of works treating such {\it non-holonomic} motion \cite{bloch2005nonholonomic} operate on the level of the equations of motion. Using well established principles, such as d'Alembert-Lagrange, Gauss or Chetaev, it is possible to construct the correct physical equations of motion \cite{flannery:2005,flannery:2011}. In turn one may determine the classical trajectory straight forwardly from the solution of these ordinary differential equations. It is important to note that these equations of motion however are not derived from an extremized action principle but instead from consideration of  forces and their relation to the variation of paths to which the system is subjected.

To date only few studies have made progress towards a genuine extremized action treatment of non-holonomic systems. Notably, in ref.~\cite{PhysRevLett.101.030402} auxiliary degrees of freedom are deployed to render the Chaplygin sleigh variational. In ref.~\cite{nonholo24} the authors use the theory of dual systems to construct a complementary system of variables that at least perturbatively allows one to treat non-holonomic systems as variational.

Our study contributes to the ongoing quest for an extremized action treatment of non-holonomic systems by independently developing an explicit and general action, based solely on the Lagrange and constraint functions. Our action captures linear and non-linear non-integrable non-holonomic motion, which remains outside of Hamilton's action principle. In contrast to \cite{PhysRevLett.101.030402} our action is formulated using the original degrees of freedom, distributed on a closed time contour and no unphysical limiting procedure is necessary. In addition, our action can be formulated non-perturbatively and its construction does not rely on the solution of a system of equations as required in \cite{nonholo24}.
 
Besides the above-mentioned systems with velocity dependent constraints the challenge of establishing an action principle affects also systems with position inequality constraints. The rolling-spinning disc on an incline \cite{flannery:2005}, as well as a particle under constant gravity in the presence of hard surfaces are two models representing motion inaccessible to Hamilton's action principle.

Non-holonomic systems are abundant in the field of robotics and autonomous transport (see e.g.\ \cite{murray2017mathematical}). Besides reliable prediction of system motion for given initial conditions (forward kinematics) inference of admissible initial data and necessary external forces to achieve a certain final state (inverse kinematics and control) is sought after (see e.g.\ \cite{baillieul2007nonholonomic}). It was shown for particular examples of holonomic motion in \cite{de_sapio_least_2008}, how an action principle in practice opens new avenues for traditional analysis and control. The application of machine-learning to robotics has led to significant improvements in their control \cite{billard2019trends}. Future progress however depends crucially on the inclusion of the mechanisms underlying robotic motion (inductive bias) into so-called physics informed neural networks \cite{kaelbling2020foundation}, information which must be provided in the form of a scalar cost functional. Non-holonomic dynamics furthermore arise in the description of contact forces \cite{hertz1881beruhrung,johnson1987contact,mindlin1949compliance} among macroscopic materials. An improved theoretical understanding of point and surface contact mechanics is key to the ongoing development of soft robots \cite{rus2015design,softrobots1} and at the same time plays a central role in reducing the energy demand of mineral and crop processing \cite{revol2001design,lisboa2007study,silveira2022study}.\\

\noindent {\bf Hamilton's principle} Hamilton showed that mechanical systems with conservative forces are variational (for a textbook see \cite{goldstein2011classical}). I.e.\ for the $d$ degrees of freedom ${\bf q}(t)\in\mathbb{R}^d$ in a system, there exists an action (cost or score) functional ${ S}=\int dt { L}[{\bf q},\dot {\bf q}]$, constructed from the system Lagrangian ${ L}$ (a measure of energy), whose critical point ${\bf q}(t)_{\rm cl}$, given boundary data ${\bf q}(t_i)_{\rm cl},{\bf q}(t_f)_{\rm cl}$, encodes the classical trajectory of the system via the following stationarity condition
\begin{linenomath*}\begin{align}
    \delta { S} = \delta\Big( \int dt L\Big)=0 \quad \Leftrightarrow \quad \frac{d}{dt}\Big(\frac{\partial { L}}{\partial \dot{q}^i}\Big)-\frac{\partial { L}}{\partial q^i}=0.\label{eq:HamVarP}
\end{align}
\end{linenomath*}
The condition on the left yields the same trajectory as the Euler-Lagrange equations of motion on the right, which are equivalent to Newton's second law. Note that at this point we have made a subtle change from a boundary value problem (left) to an initial value problem (right).

The variational formulation of classical mechanics puts symmetries center stage, since the Lagrangian transforms as a scalar under space-time and internal symmetries. Conceptually the variational formulation forms the basis for Noether's celebrated theorem \cite{noether1918invariante}, connecting symmetries and conserved charges, and in practice allows for the development of reliable numerical solvers, such as Finite Element Ritz-Galerkin methods (see e.g. \cite{galerkin2012,thomee2007galerkin}).

\noindent {\bf Constraint Forces} Constraints represent the elegant implementation of (idealized) forces ${\bf F}_{\rm C}$, acting on or within a mechanical system. Equality constraints are concisely described by constraint functions ${\bf g}({\bf q},\dot {\bf q})\in \mathbb{R}^{n\leq d}$ as ${ g}^a({\bf q},\dot {\bf q})=0$. Expressed as forces, constraints fall into the purview of Newton's second law. And while the application of Newton's equation of motion will yield the correct trajectory in principle, explicit knowledge of constraint forces prior to solving for the motion is usually not possible, preventing the application of Newton's second law in practice.

Holonomic constraints, which only depend on position, ${g}^a({\bf q})=0$ can be incorporated into Hamilton's action principle \cref{eq:HamVarP} by adjoining the constraint functions to the Lagrangian via Lagrange multiplier functions ${\bm \lambda}(t)\in\mathbb{R}^n$. The action of this \textit{generalized Hamilton's principle} then reads 
\begin{align}
    {\tilde S}=\int dt \big( { L}[{\bf q},\dot {\bf q}]+\lambda^a g^a({\bf q}) \big), \qquad \delta {\tilde S}=0
\end{align}
and one treats the multipliers as independent degrees of freedom. Similarly adjoining general non-holonomic constraints ${g}^a({\bf q},\dot {\bf q})=0$ via Lagrange multipliers fails spectacularly to produce the correct classical trajectory \cite{flannery:2011},  since the velocity dependent restrictions on the variations of the path are not implemented consistently via \cref{eq:HamVarP}.

Prior to this work, general non-holonomic constrained systems could only be treated on the level of their equations of motion. According to the Lagrange-d'Alembert principle (see \cite{flannery:2011} for a detailed derivation), which requires the constraint forces do no virtual work ${ F}_{\rm C}^i \delta {q}^i=0$, one may adjoin the equations of motion of the unconstrained system via Lagrange multipliers
\begin{linenomath*}\begin{align}
    \frac{d}{dt}\Big(\frac{\partial { L}}{\partial \dot{q}^i}\Big)-\frac{\partial { L}}{\partial q^i}={\lambda}^a\frac{\partial g^a}{\partial \dot { q}^i }\;.\label{eq:DLeom}
\end{align}\end{linenomath*}
The non-trivial $\bf \dot q$ derivative in the $\bm \lambda$ dependent term arises from the need to consider only variations consistent with the constraints and can be derived via Gauss' principle \cite{flannery:2011,papa2014}. Note that solving \cref{eq:DLeom} may lead to ${\bm \lambda}(t)$'s that are non-smooth while the physical trajectories remain smooth. \Cref{eq:DLeom} goes beyond the conventional Dirac \cite{dirac2001lectures} and Gotay-Nester \cite{AIHPA_1979__30_2_129_0} algorithms, which fail for non-integrable semi-holonomic constraints. For their recent non-holonomic generalization via the Flannery bracket see \cite{horowitzrothkopf2024}. A redefinition of velocities, motivated by the path integral (see, e.g., \cite{Jackiw:1993in}) fails when velocities enter quadratically in the constraint. \Cref{eq:DLeom} covers holonomic constraints, which by time differentiation turn into linear velocity constraints $\frac{d}{dt}g^a=\frac{\partial g^a}{\partial q^i}\dot q^i$. Information about absolute values encoded in the original holonomic constraint are provided by initial conditions. Hamilton's variational principle is unable to yield \cref{eq:DLeom}, as shown in detail in Ref.~\cite{flannery:2011}. 

Another category of non-holonomic motion evading Hamilton's principle arises from inequality position constraints $g^a({\bf q})\leq0$. They encode idealized normal (contact) forces that can lead to non-smooth trajectories. One way of treating such systems (see, e.g., \cite{Fetecau:2003}) is by following the unconstrained equations of motion in the interior of the allowed domain and manually identifying the jumps in the canonical momenta at the boundary (within the normal cone), implemented under an energy conservation constraint. Ref. \cite{alart1991mixed} on the other hand proposes a variational approach on the level of the system Lagrangian, which considers such motion as a sequence of equilibrium problems.

A causal action principle intrinsically maintains conservation laws and in addition would provide access to the trajectory globally, rendering superfluous the need to manually identify points of contact. Normal forces are closely related to sliding friction forces, such as in Coulomb friction \cite{coulomb1773sun}, which is abundant in realistic mechanical systems, and cannot be captured by either the Lagrange-d'Alembert nor Hamilton's principle.

In this work we take inspiration from an action principle originally developed in the context of quantum theory and heuristically construct from its classical limit an action for non-holonomic systems. For general velocity dependent constraints this action principle reproduces the correct Lagrange-d'Alembert equations of motion \cref{eq:DLeom} at its critical point. We also show how this action principle can be used to approximate both the non-smooth dynamics from inequality constraints and that of sliding friction on hard surfaces. In case of non-smooth trajectories our approach automatically identifies and incorporates the points of impact along the trajectory.

\section{A variational formulation of non-holonomic motion}
\label{sec2}

The need for a more general action principle beyond Hamilton is evident from the fact that in contrast to Newton's second law, \cref{eq:HamVarP} cannot capture dissipative systems with velocity dependent forces (see, e.g., \cite{goldstein2011classical}) nor velocity dependent non-holonomic constraints (see \cite{flannery:2011}). First and foremost however, Hamilton's principle is unable to capture the causal dynamics of initial value problems to start with. Indeed the need to provide boundary data ${\bf q}(t_f)_{\rm cl}$ at final time prevents the direct application of $\delta { S}=0$ to determine particle motion.

In quantum field theory it has long been known how to treat initial value problems variationally on the level of the system action via the Schwinger-Keldysh \cite{keldysh2024diagram} (or Kadanoff-Baym \cite{kadanoff1962quantum}) in-in formalism, which relies on a doubling of the degrees of freedom. In \cite{Galley:2012hx} Galley rediscovered the classical limit of the in-in formalism for point particle motion independently and understood it as a double shooting method. (For an application to classical field theory see \cite{Rothkopf:2024hxi}).

After a brief introduction to this action principle for initial value problems, we develop a a \textit{generalized classical Schwinger-Keldysh-Galley (SKG) action principle}, which produces the correct equation of motion \cref{eq:DLeom} for general velocity dependent constraints. Subsequently we use the versatility of the approach to implement the explicit constraint normal forces underlying inequality position constraints and sliding friction forces.\\

\noindent \textbf{Classical Schwinger-Keldysh-Galley Action Principle} Hamiltons's principle requires specification of acausal data ${\bf q}(t_f)_{\rm cl}$ to avoid boundary terms. As described in detail in \cite{Galley:2012hx}, classical SKG instead avoids these terms by introducing a set of doubled degrees of freedom ${\bf q}_1,\dot {\bf q}_1$ on the so-called forward time branch and ${\bf q}_2,\dot{\bf q}_2$ on the backward branch. The most general action can be written as 
\begin{linenomath*}\begin{align}
    \hspace{-0.3cm}S_{\rm SKG}=\int dt \, { L}[{\bf q}_1,\dot {\bf q}_1] - { L}[{\bf q}_2,\dot {\bf q}_2]+\Lambda[{\bf q}_1,\dot {\bf q}_2,{\bf q}_2,\dot {\bf q}_2].\label{eq:SKaction}
\end{align}\end{linenomath*}
For systems with conservative forces $\Lambda=0$, while for dissipative systems in general the action does not decompose into individual Lagrangians on each branch and $\Lambda\neq0$ takes on the role of a classical Feynman-Vernon influence functional \cite{FEYNMAN1963118}.

The variational principle is most lucidly stated after introducing the transformed coordinates ${\bf q}_+=({\bf q}_1+{\bf q}_2)/2$ and ${\bf q}_-=({\bf q}_1-{\bf q}_2)$ and reads
\begin{linenomath*}\begin{align}
 \left.\delta { S}_{\rm SKG}\right|_{{\bf q}_-=0}=0 \quad \Leftrightarrow \quad  \left.\frac{\delta { S}_{\rm SKG}}{\delta {q}_-^i}\right|_{{\bf q}_-=0,{\bf q}_+={\bf q}_{\rm cl}}=0 \;.\label{eq:SKstat}
\end{align}\end{linenomath*}
In the quantum Schwinger-Keldysh approach, ${\bf q}_+,\,{\bf q}_-$ are also known as the classical and quantum degrees of freedom, respectively \cite{altland2010condensed}. Note that by enforcing the so-called \textit{physical limit} ${\bf q}_-=0$ \textit{after} variation, the artificial doubling of the degrees of freedom is undone and we remain with the correct equations of motion for the classical trajectory ${\bf q}_+={\bf q}_{\rm cl}$.
In order to avoid acausal boundary terms we need to specify besides initial data also so-called connecting conditions
\begin{linenomath*}\begin{align}
    \nonumber &{\bf q}_+(t_i)={\bf q}_{\rm cl}(t_i), && {\bf q}_-(t_f)=0,\\
    &\underbracket{\dot {\bf q}_+(t_i)=\dot {\bf q}_{\rm cl}(t_i)}_{\rm init.\,cond.}, && \underbracket{\dot {\bf q}_-(t_f)=0}_{\rm conn.\,cond.}.\label{eq:initconcond}
\end{align}\end{linenomath*}

As Galley convincingly argues in a purely classical setting \cite{Galley:2012hx}, the doubled degrees of freedom realization of a classical action constitutes the natural extension of Hamilton's variational principle to initial value problems, as it requires no new artificial auxiliary variables and the physical limit provides a well-motivated natural procedure (associated with the $\hbar\to0$ limit) to remove the intermediate doubling.

It is important to note that the SKG action principle allows us to implement general forces ${\bf F}({\bf q},\dot {\bf q})$ via terms of the form $\Lambda= {F}^i({\bf q}_+,\dot {\bf q}_+) \,{q}_-^i$. One may thus also include dissipative, velocity dependent forces \cite{Galley:2012hx} ( for an invariant formulation in terms of retractions see e.g.~\cite{de2018variational}).  I.e.\ if an explicit form of the forces acting in a system is known, their effect can hence be straightforwardly incorporated in ${ S}_{\rm SKG}$. In the case of non-holonomically constrained systems however the key challenge is that the explicit form of the constraint forces is not known apriori. In this study we present an innovative use of doubled Lagrange multiplier degrees of freedom in the context of the SKG formalism to tackle this problem.\\

\noindent \textbf{Action for general non-holonomic systems:} The central innovation presented in this paper is a \textit{generalized classical Schwinger-Keldysh-Galley action principle} that yields the Lagrange-d'Alembert equations of motion\ in \cref{eq:DLeom} in the presence of general velocity dependent constraints described by the $n$ functions ${g}^a({\bf q},\dot {\bf q})=0$. When introducing correspondingly $n$ Lagrange multiplier functions ${\bm \lambda}$ one elevates the constraint equations to additional equations of motion of the system and deals with ${\bm \lambda}_+$ and ${\bm \lambda}_-$ as dynamical degrees of freedom. I.e.\ we must add to the classical SKG action terms that not only reproduce \cref{eq:DLeom} but that also determine the values of ${\bm \lambda}_{\rm cl}$. This can be achieved via the following \textit{generalized Schwinger-Keldysh-Galley action}
 \begin{linenomath*}\begin{align}
     {\tilde S}_{\rm SKG}\equiv\int dt \Big\{  \, { L}[{\bf q}_1,&\dot {\bf q}_1] - { L}[{\bf q}_2,\dot {\bf q}_2]\label{eq:SKactionConstr}\\
     \nonumber+ & \underbracket{{ \lambda}_-^a{ g}^a({\bf q}_+,\dot {\bf q}_+) - { \lambda}_+^a q_-^i \left. \frac{\partial { g}^a}{\partial \dot q^i}\right|_{{\bf q}={\bf q}_+}}_{\Lambda( {\bf q}_\pm, \dot {\bf q}_\pm, {\bm \lambda}_\pm)} \Big\},
\end{align}\end{linenomath*}
Variation of \cref{eq:SKactionConstr} produces the correct equations of motion \cref{eq:DLeom}. Variation w.r.t.\ ${\lambda}_-^a$ recovers ${g}^a({\bf q}_+,\dot {\bf q}_+) =0$, while the equations of motion for ${\bf q}_+$ include the correct constraint force of \cref{eq:DLeom} via the ${\lambda}_+^a$ term. These forces of constraint are often referred to as Chetaev forces \cite{chetaev1941modification}.  Note importantly that the variations $\delta q_\pm$, $\delta \dot q_\pm$, and $\delta \lambda^a_\pm$ arising from $\delta {\tilde S}_{\rm SKG}$ remain \textit{unrestricted} and the constraint functions ${g}^a({\bf q}_+,\dot {\bf q}_+)$ here are not limited to a linear dependence on the velocities. While \cref{eq:DLeom} allowed us to solve for the trajectory of general non-holonomic systems before, our work for the first time establishes how to do so from the critical point of an action.\\

As the new action \cref{eq:SKactionConstr} builds upon the well-established SKG formalism, various results follow straight-forwardly. When it comes to Noether's theorem (see e.g. \cite{sieberer_keldysh_2016} for a detailed review and the appendix of \cite{Rothkopf:2022zfb}) the doubled degrees of freedom ask us to consider two types of continuous transformations in order to establish the existence of a conserved Noether charge. We may either apply the transformation $T(\delta \alpha)$ both on the forward and backward branch $T(\delta \alpha) f_{1,2}(
\alpha) =  f_{1,2}(
\alpha+\delta \alpha)$, or instead opposite transforms on the two branches $T(\delta \alpha) f_1(
\alpha) =  f_1(
\alpha+\delta \alpha)$, $T^{-1}(\delta \alpha) f_2(
\alpha) =  f_2( 
\alpha-\delta \alpha)$. 

We may think of Noether's theorem as describing the behavior of boundary terms induced by the above mentioned transformations 
\begin{align}
\nonumber \delta S_{\rm SKG}&= \int \, dt\, \delta L_{\rm SKG}= \int dt\, \Big( \ldots {\rm e.o.m.} \ldots \Big)\label{eq:SKGNoether}\\ 
 &+ \frac{d}{dt}\Big( \underbracket{ \frac{\partial { (L+\Lambda)}}{\partial \dot{q}^i_+}\delta q^i_+}_{Q_+}  \Big) + \frac{d}{dt}\Big( \underbracket{\frac{\partial { (L+\Lambda)}}{\partial \dot{q}^i_-}\delta q^i_- }_{Q_-}\Big).
\end{align}
In the second line it is the remnants of integration by parts that form the boundary terms  $Q_+$ or $Q_-$. In the physical limit only $Q_-$ can take on finite values, as $Q_+$ contains explicit powers of $\dot q_-$. $Q_-$ therefore encodes the physical Noether charge in the SKG formalism. Note that the new $\Lambda$ term in \cref{eq:SKactionConstr} does not lead to additional contributions on the right hand side of the equality in \cref{eq:SKGNoether}. $\Lambda$ contains explicit powers of the minus d.o.f. but not their velocities; thus it does not contribute in neither $Q_+$ nor $Q_-$, in the physical limit.

Whether the Noether charge is conserved depends crucially on the transformation properties of the SKG Lagrangian. If the $L_{\rm SKG}$ is manifestly invariant then the difference of $Q_-$ at initial and final time equals zero, we consider it conserved. From the invariance under equal and opposite transforms one establishes that $\frac{d Q_+}{dt}=0$ and $\frac{d Q_-}{dt}=0$. These two results translate back into $\frac{d}{dt} \big( Q_1 - Q_2 \big) =0$ and $\frac{d}{dt} \big( Q_1+Q_2\big) =0$, respectively, and thus $\frac{d Q_1}{dt}=0$ and $\frac{d Q_2}{dt}=0$.

In case that the Lagrangian is not manifestly invariant but changes as a total derivative (such as in the case of time translations in conservative systems) we may absorb the additional boundary term into a redefinition of the $Q_-$ and consider this quantity conserved. On the other hand, if the Lagrangian transforms in an arbitrary way $Q_-$ is not conserved and $\delta L_{\rm SKG}$ encodes how $Q_-$ changes from initial to final time.

Let us consider time translations as explicit example. In conservative systems $L_{\rm SKG}$ contains terms where either $q_{\pm}$ or $\dot q_{\pm}$ appear together, but no terms that mix position and velocity of the two types. Such terms indeed transform as a total derivative under time translations. On the other hand if $L_{\rm SKG}$ contains a constraint force or dissipative force that does work, it will enter in the form of $q_- F_{\rm diss}[\dot q_+]$ and thus mix position and velocities of the plus and minus d.o.f.. In that case these terms do not transform as a total derivative and instead lead to terms such as $\dot q_+ F_{\rm diss}[\dot q_+]$ in the physical limit, which describe exactly the power dissipated by that force, which changes the associated Noether charge.

In the following we will encounter a system with a constraint quadratic in the velocities, whose energy is conserved. This may at first sight appear puzzling since terms of the form $\dot q^i_+F[\dot q^i_+]$ arise in the variation of \cref{eq:SKactionConstr} in that case. However since the resulting force term, linear in the velocity, is multiplied with one power of the same velocity and the sum over all coordinates is taken, one simply recovers the constraint equation that vanishes identically, leaving the Noether charge conserved.

\noindent \textbf{Position inequality constraints} Position inequality constraints ${g}^a(\bf q)\leq 0$ implicitly define idealized potential barriers of sufficient height, at which the propagating degrees of freedom reflect. The reversal of the momentum of the particle normal to the surface appears to happens instantaneously, leading to non-smooth trajectories. We may model such behavior explicitly with simple step functions as potentials $V^a({\bf q})=\Theta(g^a(\bf q))$, whose arguments prescribe the accessible domain. The associated normal forces ${\bf F}_N^a$ given by $({\bf F}_N^a)_i=-\delta(g^a({\bf q}))\frac{\partial g^a}{\partial {q^i}}$ are represented by a Dirac delta impulse occurring at the instant that the system makes contact with the boundary of the allowed domain. 

Sliding friction forces ${\bf F}_R^a=\mu^a |{\bf F}_N^a|(-\dot{\bf q}_{||}/|\dot {\bf q}_{||}|)$ associated with normal forces ${\bf F}_N^a$ via the respective coefficient of kinetic friction $\mu^a$, on the other hand, oppose motion and thus acquire a dependence on velocity along the contact surface.

In the classical SKG approach we can incorporate the physics of contact and sliding friction directly on the level of a generalized interaction term as
\begin{linenomath*}\begin{align}
     &S_{\rm SKG}=\label{eq:SKactionIneq}\int dt \, { L}[{\bf q}_1,\dot {\bf q}_1]- { L}[{\bf q}_2,\dot {\bf q}_2] \\
     \nonumber & + {q}_-^i\left.\sum_a ({\bf F}^a_{N})_i\right|_{{\bf q}={\bf q}_+}  + q^i_- \left.\sum_a \mu^a |{\bf F}_N^a| \frac{-\dot q^i_{||}}{|\dot {\bf q}_{||}|}\right|_{{\bf q}={\bf q}_+},
\end{align}\end{linenomath*}
where $\dot {\bf q}_{||}$ denotes the projection of the velocity parallel to the contact surface. Variation of \cref{eq:SKactionIneq} produces the correct global, i.e. not necessarily smooth, trajectory, inaccessible locally via \cref{eq:DLeom}.

Note that the normal force term does not depend on the velocities $\dot {\bm q}$ and thus does not affect the invariance of the action under time translations, guaranteeing the preservation of energy by construction. While the constraint potential $V^a({\bf q})$ could be included in Hamilton's principle, only the classical Schwinger-Keldysh-Galley approach allows one to formulate this physical setup as an initial value problem. The sliding friction force does depend on velocity, exposing its dissipative character.

Since contact is not actually instantaneous in nature, we may regularize the associated normal force delta impulse by a Gaussian with a width $\sigma$. This width $\sigma$ is chosen small enough to appear localized within the desired level of accuracy.

\section{Application to model systems}

\begin{figure}
    \centering
    \includegraphics[scale=0.82]{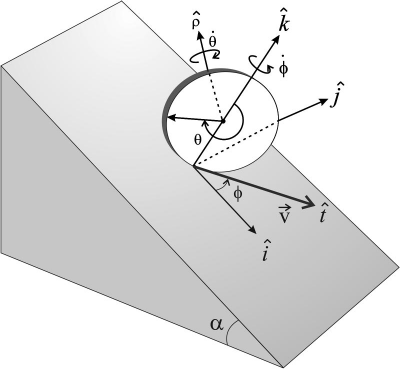}
    \caption{{\bf Sketch of the rolling-spinning disk on an incline.} $x,y$ encode the center of mass ${\bf r}_{\rm c.m.}=x\hat i + y\hat j+z\hat k$ and the angles $\theta,\phi$ encode rolling and spinning motion. Reprinted from \cite{flannery:2005}, with the permission of AIP Publishing.}
    \label{fig:DiskSketch}
\end{figure}

\noindent \textbf{Non-holonomic velocity constraints} Our first application is the rolling-spinning disk \cite{flannery:2005} with radius $R$ and mass $M$ on an incline of angle $\alpha$ in the presence of constant gravity $g$, sketched in \cref{fig:DiskSketch}. The system is described by four interdependent degrees of freedom $x,y,\theta$ and $\phi$. The former two encode the position of the center of mass ${\bf r}_{\rm c.m.}=x\hat i + y\hat j+z\hat k$, while the latter two angles $\theta,\phi$ encode the rolling and spinning motion. Given the moment of inertia $I_S$ about the symmetry axis and $I_D$ about the fixed spinning axis, the Lagrangian reads
\begin{linenomath*}\begin{align}
    \hspace{-0.25cm}{ L}_{\rm D}=\frac{1}{2}M(\dot x^2+\dot y^2)+\frac{1}{2}I_S\dot \theta^2 + \frac{1}{2}I_D\dot\phi^2+Mgx{\rm sin}(\alpha).\label{eq:Ldisk}
\end{align}\end{linenomath*}
For non-slip motion, we ensure $g_{\rm D1}=\dot x^2+\dot y^2-R^2\dot\theta^2=0$ and $g_{\rm D2}=\dot x {\rm sin}(\phi) - \dot y {\rm cos}(\phi)=0$. Both constraints are non-integrable but captured by the Lagrange-d'Alembert principle of \cref{eq:DLeom}.

\begin{figure}
    \centering
    \includegraphics[scale=0.34]{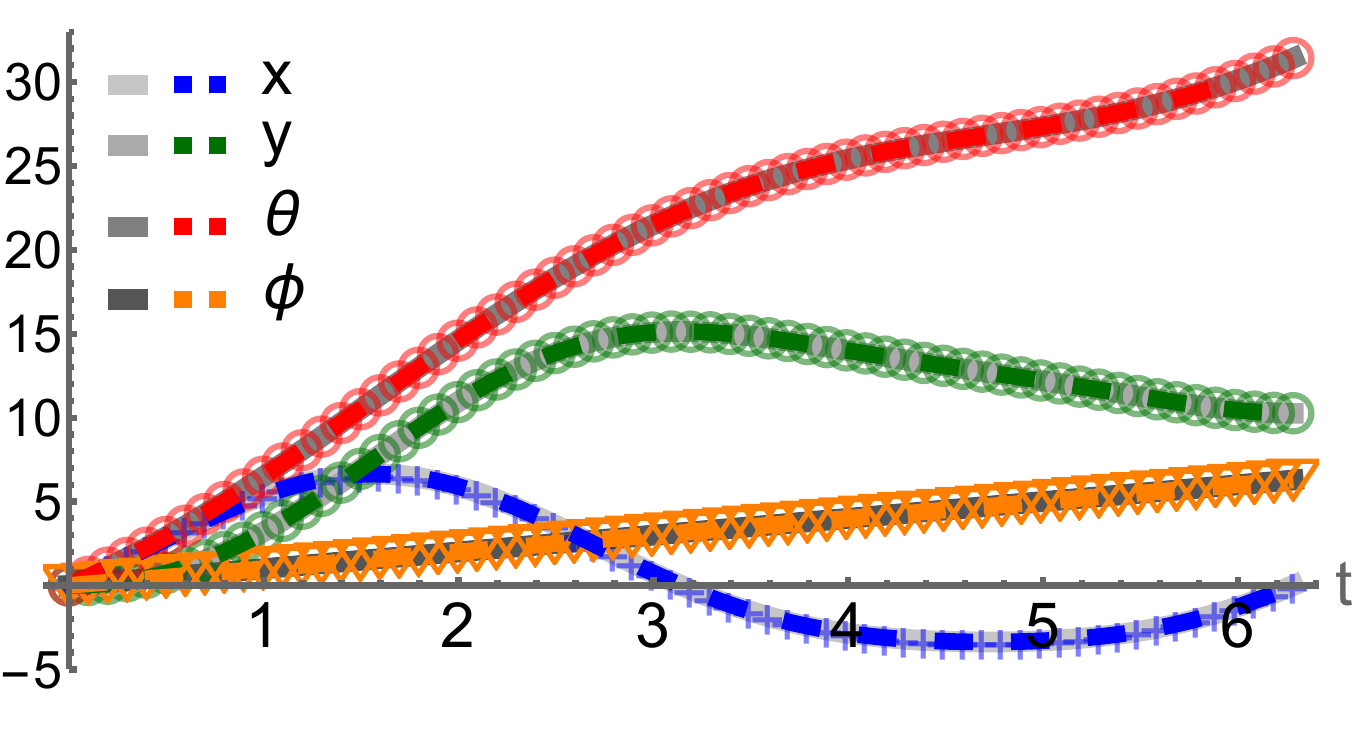}
    \includegraphics[scale=0.4]{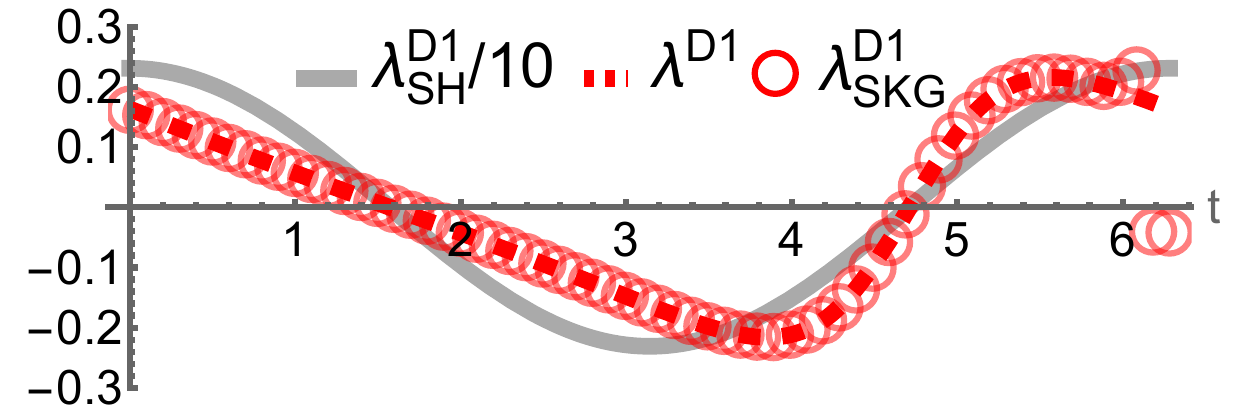}
        \includegraphics[scale=0.4]{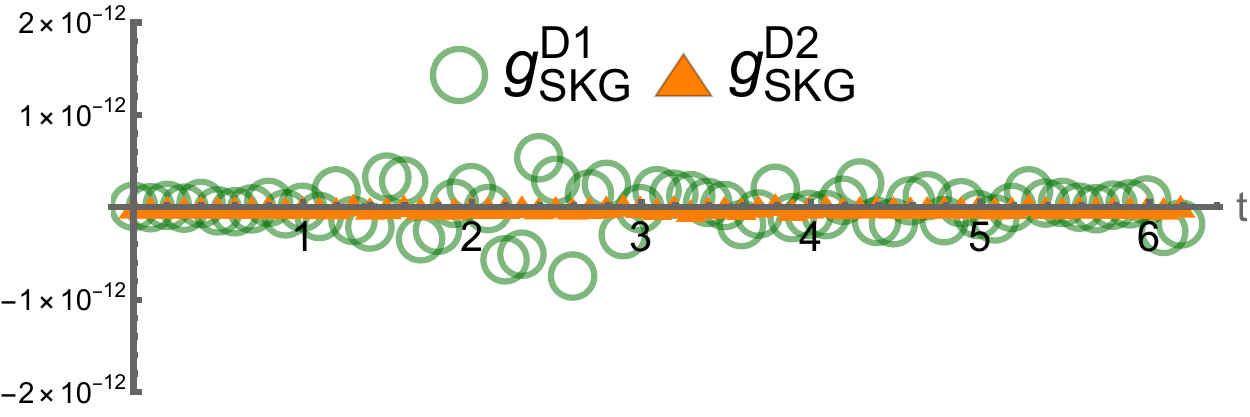}
               \includegraphics[scale=0.4]{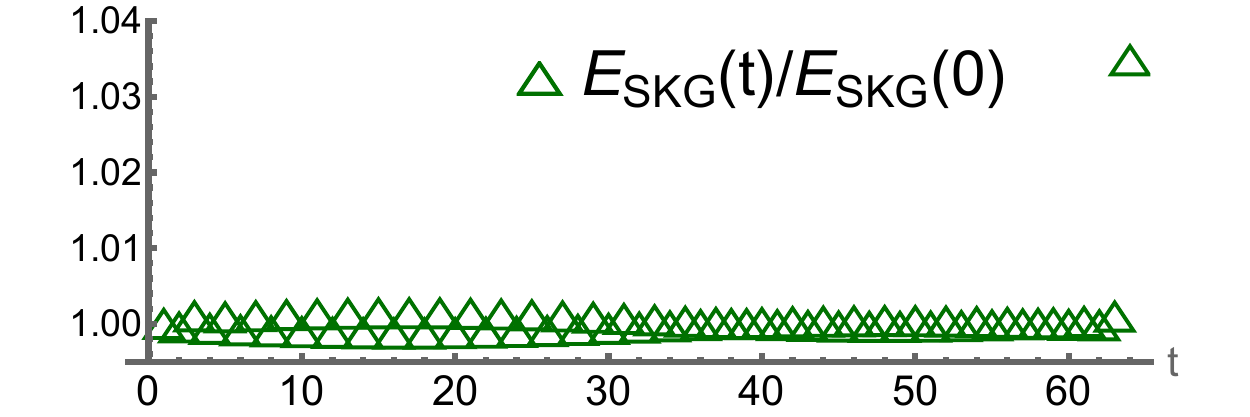}
            \caption{{\bf Time evolution of a rolling-spinning disc}. Motion obtained via the semi-holonomic (solid gray) and non-holonomic (dashed color) Lagrange-d'Alembert equations of motion as well as via the critical point of our novel classical SKG action \cref{eq:discrSvc} (color symbols). Physical degrees of freedom are shown in the top-, the Lagrange multiplier to $g^{\rm D1}$ in the second from top panel. We evaluate the constraint functions $g^{\rm D1}$ and $g^{\rm D2}$ on the solution from our action in the second to bottom panel and the energy in the bottom panel. Due to the presence of the Lagrange multipliers for the connecting condition at the last time slice, one observes benign (diminishing under grid refinement) deviations at the last and second to last time slice. For $\lambda^{\rm D1}_{\rm SKG}$ and the energy close to $t_f$ the last two points are affected, while for the constraints only the last time step is affected.}
    \label{fig:DiskEvol}
\end{figure}

With the help of $g_{\rm D2}$, one can turn $g_{\rm D1}$ into semi-holonomic form \cite{flannery:2005}, for which we solve \cref{eq:DLeom} using the \texttt{Mathematica} \texttt{NDSolve} command \cite{Mathematica}.  We plot these results\footnote{All results shown in \protect\cref{fig:DiskEvol} use $g=9.8$ ${\rm m/s^2}$, $\alpha=\pi/6$, $M=1$ kg, $R=1$ m, $I_S=1/2$ kg ${\rm m^2}$ and $I_D=1/4$ kg ${\rm m^2}$, as well as initial values $x_i=0$ m, $\dot x_i=5$ ${\rm m/s}$ , $y_i=0$ m, $\dot y_i=0$ ${\rm m/s}$, $\theta_i=0$, $\dot \theta_i=5$ rad/s, $\phi_i=0$, $\dot \phi_i=1$ rad/s.} as gray solid lines in \cref{fig:DiskEvol}. Physical trajectories are shown in the top panel and the Lagrange multiplier $\lambda^{\rm D1}_{\rm SH}$ associated with the linearized semi-holonomic $g_{\rm D1}$ in the bottom panel. If we instead solve \cref{eq:DLeom} with the non-linear $g_{\rm D1}$ we obtain the colored dashed lines. Note that the same physical trajectories are obtained, while $\lambda^{\rm D1}$, now enforcing another constraint, differs\footnote{A reference implementation of all numerical examples is available at \cite{rothkopfZ:2024}.}.

Using \cref{eq:Ldisk} with the non-linear $g_{\rm D1}$ and $g_{\rm D2}$ in \cref{eq:SKactionConstr}, introducing $\lambda^{\rm D1}_{1,2}$ and $\lambda^{\rm D2}_{1,2}$ Lagrange multiplier functions to encode the two constraints, one obtains the corresponding classical Schwinger-Keldysh-Galley action.

While the construction of the generalized SKG action \cref{eq:SKactionConstr} and the action \cref{eq:SKactionIneq} constitutes the central conceptual contribution of this work, we believe that numerical application will provide the reader with additional evidence for its veracity. Hence we present below an independently developed strategy to solve for the classical trajectory numerically in its full non-linear non-holonomic form on the action level, bypassing governing equations. Our approach on the action level is complementary to the variational solvers for non-holonomic system presented in the literature (see e.g. \cite{Marsden_West_2001, cortes2001non,ferraro2008momentum,mclachlan2006integrators,10.1016/j.cam.2022.114837,modin2020makes}), which are formulated in terms of governing equations.

To numerically determine the trajectory we evaluate the critical point of this action after discretization, following \cite{Rothkopf:2022zfb}. We rely on summation-by-parts (SBP) finite difference operators (for reviews see \cite{svard2014review,fernandez2014review,lundquist2014sbp}), which mimic integration-by-parts (IBP) exactly in the discrete setting and deploy a time grid with $N$ steps $\Delta t=(t_f-t_i)/(N-1)$. Details of the discretization procedure and the explicit form of the numerical action is provided in \cref{sec:appdiscrI}. 

Choosing a fourth order accurate SBP discretization \texttt{SBP242} on $N=64$ points to span the time interval $t_f-t_i=2\pi$ s, we carry out a numerical optimization with the \texttt{IPOTP} library accessible through the \texttt{FindMinimum} command of \texttt{Mathematica} to obtain the critical point of the discretized action \cref{eq:discrSvc}. The result is plotted as colored symbols in \cref{fig:DiskEvol} and shows excellent agreement with the solution of \cref{eq:DLeom} given as dashed colored lines.

Note that the non-linear velocity constraint, as well as the normal force constraint are excellently conserved within the simulated time interval (second to bottom panel). Similarly the energy of the system (bottom panel) shows only minute deviations from its initial value within the simulation domain.\\
Due to the presence of Lagrange multipliers enforcing the connecting condition, we must expect deviations to arise on the temporal boundary. These artifacts are benign, as they diminish with grid refinement and do not spoil the convergence of the discretization, as studied in more detail in \cite{Rothkopf:2022zfb}. And indeed, we find that $\lambda^{\rm D1}_{\rm SKG}$, as well as the energy show a deviation from their  smooth behavior at the last two time steps. The constraint functions too show a deviation at $t_f$, which lies outside the window of $10^{-12}$ chosen here to achieve good visibility of the differences between $g^{\rm D1}$ and $g^{\rm D2}$.\\

\begin{figure}[t]
    \centering
    \includegraphics[scale=0.25]{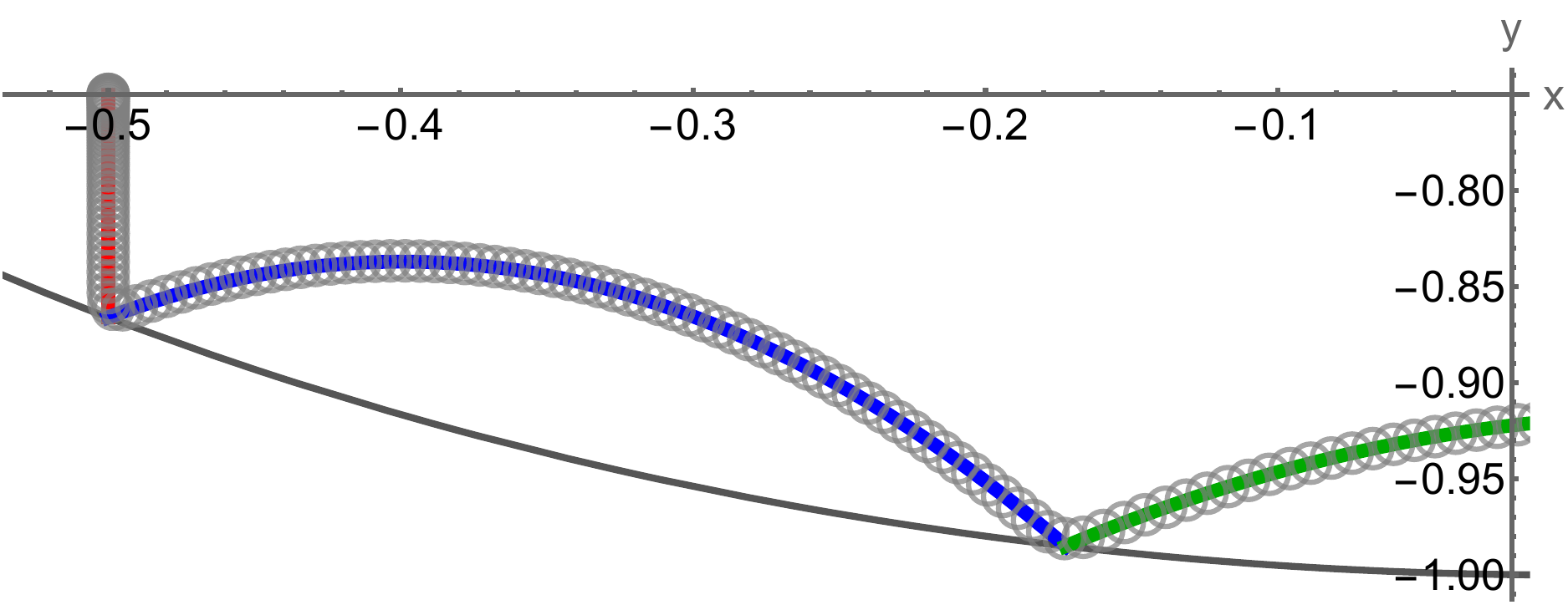}
    \includegraphics[scale=0.25]{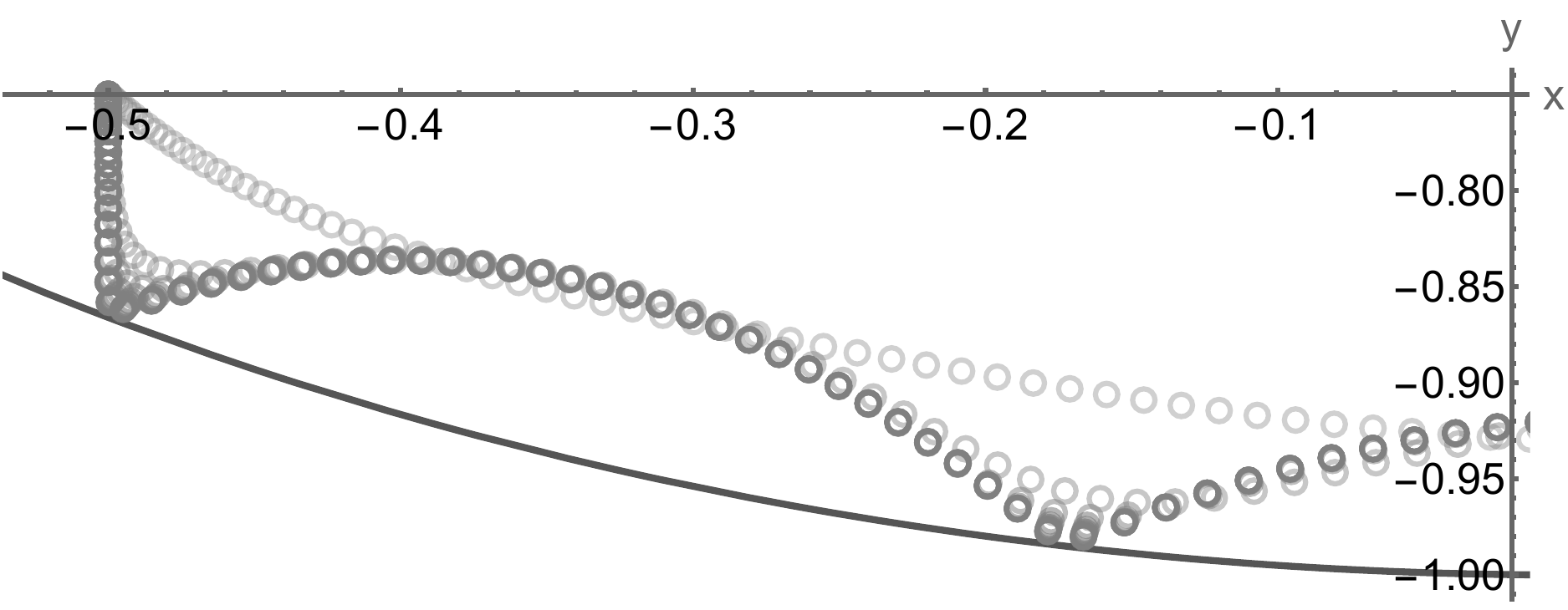}
    
    \vspace{-0.3cm}
    \includegraphics[scale=0.32]{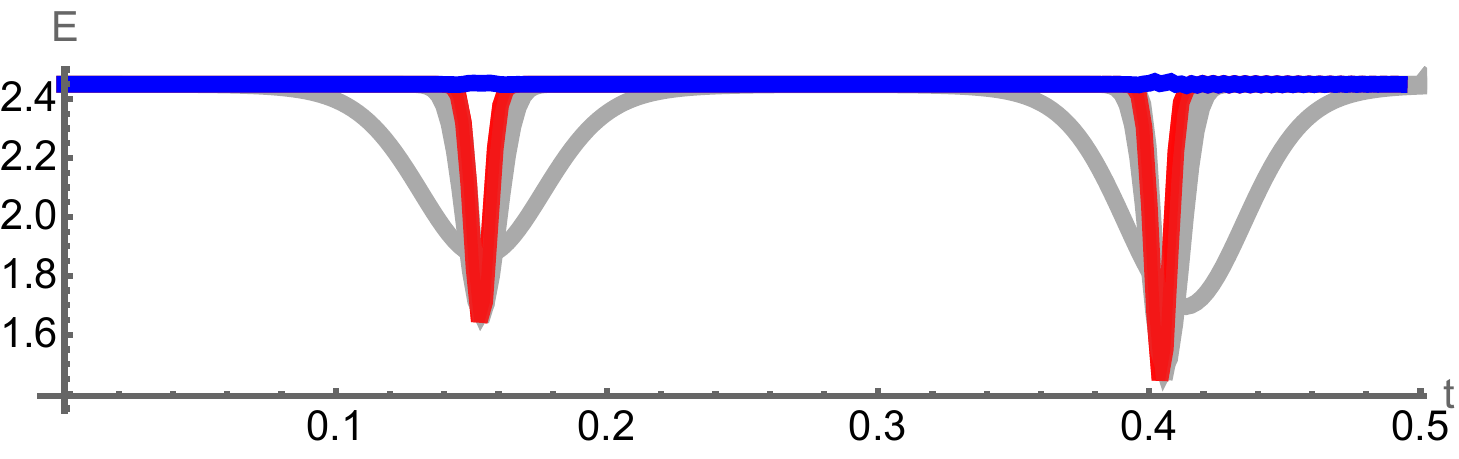}

    \caption{{\bf Point mass falling in a tumbler}. (Top) Motion obtained via Newton's 2nd law (solid colors) and from the critical point of our \texttt{SBP424} discretized action (gray circles) with $N=256$ and $\sigma=1/110$. (Center) Convergence of the critical point towards the correct solution as the width of the Gaussian $\sigma=1/10\ldots1/110$m is decreased (light to dark gray). (Bottom) Total energy $E$ is shown as blue solid curve. Its kinetic and gravitational contribution are given in red. The gray lines show the change in kinetic and gravitational contributions as $\sigma$ is reduced.}
    \label{fig:bouncemotion}
\end{figure}

\noindent \textbf{Position inequality constraints} Our second application is the particle under constant gravity in a hard-walled tumbler in the absence of friction. The Lagrangian of the unconstrained system is 
\begin{linenomath*}\begin{align}
    { L}_{\rm S}=\frac{1}{2}m(\dot x^2+ \dot y^2) - mgy.
    \label{eq:2dgravitylagrangian}
\end{align} \end{linenomath*}

The accessible regime is given by $g_{\rm S}= x^2+y^2-R^2 \leq0$, which we implement by the step function potential $V(x,y)=V_0\Theta(g_{\rm S})$. We regulate the resulting delta impulse force on the boundary in \cref{eq:SKactionIneq} using a Gaussian of width $\sigma$, which leads us to ${\bf F}_N({\bf r})= -\frac{V_0}{\sqrt{2\pi\sigma^2}}e^{-(x^2+y^2-R^2)^2/2\sigma^2}{\bf r}$ where ${\bf r}=(x,y)$. In order for the particle to be reflected we must choose $V_0$ larger than the largest possible kinetic energy the particle may acquire. That value is provided by the conserved initial energy of the system. 

\begin{figure}[t]
    \centering
    \includegraphics[scale=0.37]{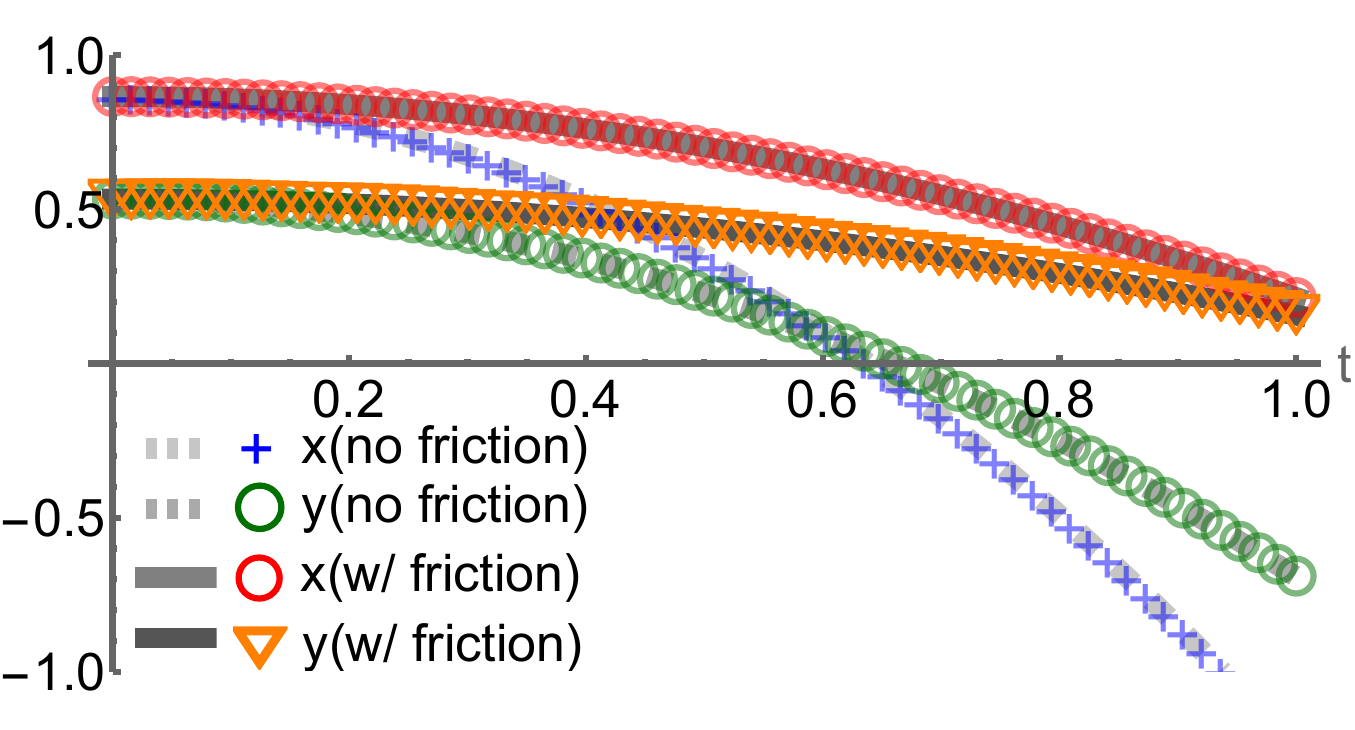}
    \caption{{\bf Sliding motion along an incline of angle $\alpha=\pi/6$ in constant gravity.} Trajectories from Newton's second law in gray in the absence (dashed) and presence (solid) of kinetic Coulomb friction $\mu=4/10$. Trajectories from the critical point of action \cref{eq:discrSF} are given as open colored symbols, where the normal force is regularized with $\sigma=1/70$m and initial positon is offset by $\delta=1/30$m.}
    \label{fig:slidingmotion}
\end{figure}

Being a system that can produce non-smooth trajectories, we set out to solve it numerically, directly from the critical point of the SBP discretized action, whose explicit expression is provided in \cref{sec:appdiscrII}.

Note that a large $\sigma/R\sim 1$ corresponds to a fuzzy boundary and only by reducing its value will one approach the hard wall limit. One can determine self-consistently the smallest admissible choice of $\sigma$ for a given time resolution by monitoring the mechanical energy $E=\frac{1}{2}M \big( (dx/dt)^2 + (dy/dt)^2 \big) + Mgy$. The discretized energy is given explicitly in \cref{sec:appdiscrII}. As the Gaussian impulse becomes too sharp to be resolved, the energy after impact with the wall will show oscillatory artifacts around its well preserved value. 

In the top panel of \cref{fig:bouncemotion} we show our numerical result for the critical point of \cref{eq:discrSvi}, discretized with the fourth order accurate \texttt{SBP424} scheme, obtained via the \texttt{IPOTP} method of \texttt{Mathematica}'s  \texttt{FindMinimum} command. The gray open symbols correspond to a point mass with $M=1$ kg in a tumbler of radius $R=1$ m, positioned initially at rest at $x_i=-1/2$ m, $y_i=-3/4$ m. The time interval is discretized with $N=256$ steps of size $\Delta t=\frac{1}{2(N-1)}$ s and the delta impulse with $V_0=10$ J regulated with a width parameter $\sigma=1/110$ m. We find excellent agreement with the manually computed trajectory according to Newton's second law, shown as solid colored lines. Note that in the action based approach we did not have to locate the points of impact manually, as the method automatically generates the complete global trajectory of the particle.

In the center panel of \cref{fig:bouncemotion} we show that our solution systematically approaches the non-smooth trajectory for the hard-wall tumbler, as the parameter $\sigma$ is reduced from $1/10$ m to $1/110$ m. The reason we stop at $\sigma=1/110$ m follows from an analysis of the total energy in the bottom panel shown as blue line. Upon magnification, an inspection by eye reveals that the energy begins to exhibit small oscillatory artifacts close to $t=1/2$ s after the second impact. We have checked that increasing $N$ and decreasing $\Delta t$ allows us to further reduce $\sigma$. The red curve in the bottom panel denotes the mechanical energy $E$ without the contributions from the boundary. The regularized delta impulse with $\sigma=1/110$ m temporarily changes the kinetic energy of the particle, but returns it to the same total mechanical energy as before the impact. The gray lines illustrate the different changes in mechanical energy for larger values of the $\sigma$ parameter.

As a final example, we demonstrate how one can incorporate normal and friction forces in our action approach. Consider the motion of a point particle mass sliding down an inclined plane of angle $\alpha=\pi/6$ whose surface has a non-zero coefficient of kinetic friction $\mu$. Since the mass cannot penetrate the surface of the incline, we have an inequality constraint $x\tan\alpha-y\ge0$ that we treat in addition to the standard Lagrangian \cref{eq:2dgravitylagrangian}. Let us introduce the normal force, regularized by a Gaussian with width $\sigma$, which reads ${\bf F}_N= -\frac{V_0}{\sqrt{2\pi\sigma^2}}e^{-(x\tan\alpha-y)^2/2\sigma^2}(\tan\alpha\hat x-\hat y)$.  The kinetic friction force then reads ${\bf F}_R= \mu \frac{V_0}{\sqrt{2\pi\sigma^2}}e^{-(x\tan\alpha-y)^2/2\sigma^2}(\hat x+\tan\alpha\hat y)$. Note that due to the regularization with $\sigma$, we also must offset the sliding mass point away from the exact position of the hard surface it moves along by a distance $\delta$. As the regularization $\sigma$ of the normal force is diminished with increasing number of grid points $N$, similar to the previous example, the offset $\delta$ too may be reduced.

In \cref{fig:slidingmotion} we present numerical results for motion on the incline from the critical point of the discretized action given explicitly in \cref{eq:discrSF} and obtained via the \texttt{IPOTP} method of \texttt{Mathematica}'s  \texttt{FindMinimum} command. We have chosen a time interval of $t_f-t_i=1$ s discretized on $N=64$ points and deploy the \texttt{SBP424} discretization scheme. Since the motion in this case is smooth $N=64$ suffices to achieve result already visually indistinguishable from Newton's law, where for non-smooth motion $N=256$ was required. Motion is initialized at $x_i=\cos\alpha$ m and $y_i=(\sin\alpha+\delta)$ m with $\delta=1/30$ m with the potential barrier set at $V_0=5$ J. The offset $\delta$ was chosen to minimize residual oscillatory artifacts arising from the regularization of the normal force with $\sigma=1/70$ m. 

Comparing in \cref{fig:slidingmotion} to the direct solution of Newton's second law in the absence (gray dashed) and presence (gray solid) of friction we find excellent agreement of the action based results (colored open symbols). We have checked that reducing $\sigma$ together with $\delta$ quantitatively improves the agreement between the two approaches.

\section{Conclusion}

In this work, we have put forward an explicit and general extremized action formulation for systems with non-holonomic constraints, drawing on the flexibility of doubled degrees-of-freedom in the Schwinger-Keldysh-Galley formalism. Our approach incorporates both holonomic and non-holonomic equality and inequality constraints.

Our work contributes to broadening the scope of classical variational mechanics and offers a novel path for future explorations of constrained quantum dynamics, such as those relevant to nanoscale machines and artificial molecular motors \cite{erbas2015artificial,PhysRevLett.101.030402,fernandez2018quantum}.

By showing that non-holonomic motion can be cast in an action form via the classical limit of the quantum Schwinger-Keldysh action formalism, our work contributes to a long-standing effort to unify classical mechanics under the principle of extremized action. We anticipate that this formulation will provide new analytical and computational tools for a wide range of applications in physics and engineering.

\begin{acknowledgments}
A.\ R.\ thanks E. Grossi and Z.\ Fodor for discussions. A.\ R.\ and W.\ A.\ H.\ acknowledge support by the ERASMUS+ project 2023-1-NO01-KA171-HED-000132068. A.\ R.\ was supported by a Korea University Grant. W.\ A.\ H.\ thanks the South African National Research Foundation and the SA-CERN Collaboration for financial support and the BNL EIC theory institute for its hospitality. This research was conducted in part by WAH while visiting the Okinawa Institute of Science and Technology (OIST) through the Theoretical Sciences Visiting Program (TSVP).
\end{acknowledgments}

\section*{Author Contributions}
A.\ R.\  put forward the idea of exploiting the doubled degrees of freedom formalism to capture quadratic velocity dependent non-holonomic constraints. A.\ R.\  and W.\ A.\ H.\ with equal contribution developed the classical SKG formalism for velocity dependent constraints, as well as the formalism to incorporate inequality constraints based on generalized forces. A.\ R.\  implemented the formalism for the example systems discussed in this paper, available as supplementary material. Editing of the manuscript by both A.\ R.\  and W.\ A.\ H.\ based on an original draft by A.\ R.\ .

\section*{Data availability statement}

Mathematica notebooks implementing all numerical examples are openly available at \cite{rothkopfZ:2024}.

\section*{Additional Information}

\noindent Correspondence and requests for materials should be addressed to Alexander Rothkopf.\\

\noindent The authors declare no competing interests.

\bibliography{VariationalApproachNonholonomic}

\appendix
\section{Methods}

\subsection{Discretized action for the rolling-falling disk}
\label{sec:appdiscrI}
Given a time grid with $N$ steps $\Delta t=(t_f-t_i)/(N-1)$, the physical degrees of freedom of the rolling-falling disk are represented by $N$-component arrays ${\bf x}_{1,2}$, ${\bf y}_{1,2}$, ${\bm \theta}_{1,2}$ and ${\bm \phi}_{1,2}$. Integration by parts connects integration and differentiation, hence we must choose a compatible quadrature matrix $\mathds{H}$ implementing $\int dt f(t) g(t) \approx {\bf f}^T \mathds{H} {\bf g}$ and finite difference operators $\mathds{D}=\mathds{H}^{-1}\mathds{Q}$. Here the matrix $\mathds{Q}$ encodes the finite difference stencil structure and ensures the SBP property via $\mathds{Q}^T+\mathds{Q}={\rm diag}[-1,0,\ldots,0,1]$. The lowest order diagonal \texttt{SBP121} approach combines the trapezoid rule $\mathds{H}=\Delta t{\rm diag}[\frac{1}{2},1,\ldots,1,\frac{1}{2}]$ with the the lowest order central symmetric finite difference stencil in the interior, and the forward and backward stencils on the backward and forward boundary respectively (for details and explicit expression for the higher order \texttt{SBP242} see \cite{Rothkopf:2022zfb}). In a numerical setting the initial conditions and connecting conditions of \cref{eq:initconcond} must be included explicitly on the action level, which we accomplish by adding additional Lagrange multiplier variables, $\kappa$ for initial and $\xi$ for connecting conditions, to the action functional \cref{eq:SKactionConstr} \footnote{Initial data provides appropriate regularization of the SBP operators as described in detail in \cite{Rothkopf:2022zfb}}. The explicit expression reads

\begin{widetext}
\begin{align}
        \nonumber \mathds{S}_{\rm D}= &\frac{1}{2}M(\mathds{D}{\bf x}_+)^T\mathds{H}(\mathds{D}{\bf x}_-)+\frac{1}{2}M(\mathds{D}{\bf y}_+)^T\mathds{H}(\mathds{D}{\bf y}_-)+\frac{1}{2}I_D(\mathds{D}{\bm \theta}_+)^T\mathds{H}(\mathds{D}{\bm \theta}_-)+\frac{1}{2}I_S(\mathds{D}{\bm \phi}_+)^T\mathds{H}(\mathds{D}{\bm \phi}_-)+Mg{\rm sin}(\alpha)\mathds{1}^T \mathds{H} {\bf x}_-\\
    &+ \kappa_x({\bf x}_+[1]-x_i) + \tilde \kappa_x( (\mathds{D}{\bf x}_+)[1]-\dot x_i)+ \xi_x ({\bf x}_-[N]) + \tilde \xi_x ({\bf x}_-[N])\label{eq:discrSvc}\\
    \nonumber &+ \kappa_y({\bf x}_+[1]-y_i) + \tilde \kappa_y( (\mathds{D}{\bf y}_+)[1]-\dot y_i)+ \xi_y ({\bf y}_-[N]) + \tilde \xi_y ({\bf y}_-[N])\\
    \nonumber &+ \kappa_\theta({\bm \theta}_+[1]-\theta_i) + \tilde \kappa_\theta( (\mathds{D}{\bm \theta}_+)[1]-\dot \theta_i)+ \xi_\theta ({\bm \theta}_-[N]) + \tilde \xi_\theta ({\bm \theta}_-[N])\\
    \nonumber &+ \kappa_\phi({\bm \phi}_+[1]-\phi_i) + \tilde \kappa_\phi( (\mathds{D}{\bm \phi}_+)[1]-\dot \phi_i)+ \xi_\phi ({\bm \phi}_-[N]) + \tilde \xi_\phi ({\bm \phi}_-[N])\\
    \nonumber &+{\bm \lambda}^{\rm D1}_{-}\circ \Big( (\mathds{D}{\bf x}_+)^T\mathds{H}(\mathds{D}{\bf x}_+ ) + (\mathds{D}{\bf y}_+)^T\mathds{H}(\mathds{D}{\bf y}_+) -R^2(\mathds{D}{\bm \theta}_+)^T\mathds{H}(\mathds{D}{\bm \theta}_+)  \Big)\\
    \nonumber &-{\bm \lambda}^{\rm D1}_{+}\circ \Big( (\mathds{D}{\bf x}_+)^T \mathds{H} {\bf x}_- + (\mathds{D}{\bf y}_+)^T \mathds{H} {\bf y}_- - R^2 (\mathds{D}{\bm \theta}_+)^T \mathds{H} {\bm \theta}_- \Big)\\
    \nonumber &+{\bm \lambda}^{\rm D2}_{-}\circ\Big( (\mathds{D}{\bf x}_+)^T\mathds{H} {\rm sin}[{\bm \phi}] + (\mathds{D}{\bf y}_+)^T\mathds{H} {\rm cos}[{\bm \phi}] \Big) -({\bm \lambda}^{\rm D2}_{+})^T \Big( \mathds{H} {\rm sin}[{\bm \phi}] + \mathds{H} {\rm cos}[{\bm \phi}]\Big) \\
    \nonumber &+ \kappa_{\lambda^{\rm D1}}( {\bm \lambda}^{\rm D1}_-[1]) + \xi_{\lambda^{\rm D1}}( {\bm \lambda}_-^{\rm D1}[N])+ \kappa_{\lambda^{\rm D2}}( {\bm \lambda}^{\rm D2}_-[1])+  \xi_{\lambda^{\rm D2}}( {\bm \lambda}^{\rm D2}_-[N])
 \end{align}
\end{widetext}
The symbol $\circ$ refers to element-wise multiplication with the array to the right. Since ${\bm \lambda}^{\rm D1}_-$ and ${\bm \lambda}^{\rm D2}_-$ appear in combination with derivative terms we need to set them zero at initial and final time via Lagrange multipliers in the last line to avoid unphysical boundary contributions.

Due to the use of a mimetic SBP discretization scheme, the discretized mechanical energy ${\bf E}$ along time is obtained from the continuum expression simply by replacing derivatives by finite difference operators $\mathds{D}$ 
\begin{widetext}\begin{align}
&{\bf E}_{\rm D}=\frac{1}{2}M(\mathds{D}{\bf x}_+)\circ(\mathds{D}{\bf x}_+)+\frac{1}{2}M(\mathds{D}{\bf y}_+)\circ(\mathds{D}{\bf y}_+)+\frac{1}{2}I_D(\mathds{D}{\bm \theta}_+)\circ(\mathds{D}{\bm \theta}_+)+\frac{1}{2}I_S(\mathds{D}{\bm \phi}_+)\circ(\mathds{D}{\bm \phi}_+)-Mg{\rm sin}(\alpha){\bf x}_+
\end{align}\end{widetext}

\subsection{Discretized action for the particle in a tumbler}
\label{sec:appdiscrII}

Here too physical degrees of freedom are represented on a time grid with $N$ steps $\Delta t=(t_f-t_i)/(N-1)$ by $N$-component arrays ${\bf x}_{1,2}$, ${\bf y}_{1,2}$. Deploying the same SBP discretization described in \cref{sec:appdiscrI}, adding appropriate Lagrange multipliers to fix the initial and connecting conditions, we obtain the following expression for the discretized action
\begin{widetext}
\begin{align}
\mathds{S}_{\rm S}= &\frac{1}{2}M(\mathds{D}{\bf x}_+)^T\mathds{H}(\mathds{D}{\bf x}_-)+\frac{1}{2}M(\mathds{D}{\bf y}_+)^T\mathds{H}(\mathds{D}{\bf y}_-)\label{eq:discrSvi} -Mg\mathds{1}^T \mathds{H} {\bf y}_-\\
\nonumber &-\frac{V_0}{\sqrt{2\pi \sigma^2}}\Big({\rm exp}\Big[-( ({\bm y}_+)^2+({\bm x}_+)^2 -1 )^2/(2\sigma^2)\Big]\Big)^T \mathds{H} \big( {\bm x}_-\circ{\bm x}_+ + {\bm y}_-\circ{\bm y}_+\big)\\
\nonumber &+ \kappa_x({\bf x}_+[1]-x_i) + \tilde \kappa_x( (\mathds{D}{\bf x}_+)[1]-\dot x_i)+ \xi_x ({\bf x}_-[N]) + \tilde \xi_x ((\mathds{D}{\bf x}_-)[N])\\
\nonumber &+ \kappa_y({\bf y}_+[1]-y_i) + \tilde \kappa_y( (\mathds{D}{\bf y}_+)[1]-\dot y_i)+ \xi_y ({\bf y}_-[N]) + \tilde \xi_y ((\mathds{D}{\bf y}_-)[N])
\end{align}
\end{widetext}
Note that raising a discrete array to a power, as well as applying the exponential function acts element-wise in the above expression. 

Due to the use of a mimetic SBP discretization scheme, the discretized mechanical energy ${\bf E}$ along time is obtained from the continuum expression simply by replacing derivatives by finite difference operators $\mathds{D}$ 
\begin{linenomath*}\begin{align}
&{\bf E}_{\rm S}=\\
\nonumber &\frac{1}{2}M(\mathds{D}{\bf x}_+)\circ(\mathds{D}{\bf x}_+)+ \frac{1}{2}M(\mathds{D}{\bf y}_+)\circ(\mathds{D}{\bf y}_+) + Mg {\bf y}_+
\end{align}\end{linenomath*}

\subsection{Discretized action for the particle sliding on an incline}
\label{sec:appdiscrIII}

Again physical degrees of freedom are represented on a time grid with $N$ steps $\Delta t=(t_f-t_i)/(N-1)$ by $N$-component arrays ${\bf x}_{1,2}$, ${\bf y}_{1,2}$. Deploying the same SBP discretization described in \cref{sec:appdiscrI} and \cref{sec:appdiscrII}, adding appropriate Lagrange multipliers to fix the initial and connecting conditions, we obtain the following expression for the discretized action
\begin{widetext}
\begin{linenomath*}\begin{align}
     \mathds{S}_{\rm SF}= &\frac{1}{2}M(\mathds{D}{\bf x}_+)^T\mathds{H}(\mathds{D}{\bf x}_-)+\frac{1}{2}M(\mathds{D}{\bf y}_+)^T\mathds{H}(\mathds{D}{\bf y}_-) -Mg\mathds{1}^T \mathds{H} {\bf y}_-\label{eq:discrSF}\\
     \nonumber &-\frac{V_0}{\sqrt{2\pi \sigma^2}}\Big({\rm exp}\Big[-(  {\rm tan}[\alpha] {\bm x}_+ - {\bm y}_+  )^2/(2\sigma^2)\Big]\Big)^T \mathds{H} \big( {\rm tan}[\alpha]{\bm x}_- - {\bm y}_-\big)\\
     \nonumber &+\mu \frac{V_0}{\sqrt{2\pi \sigma^2}}\Big({\rm exp}\Big[-(  {\rm tan}[\alpha] {\bm x}_+ - {\bm y}_+  )^2/(2\sigma^2)\Big]\Big)^T \mathds{H} \big( {\bm x}_- + {\rm tan}[\alpha]{\bm y}_-\big)\\
     \nonumber &+ \kappa_x({\bf x}_+[1]-x_i) + \tilde \kappa_x( (\mathds{D}{\bf x}_+)[1]-\dot x_i)+ \xi_x ({\bf x}_-[N]) + \tilde \xi_x ((\mathds{D}{\bf x}_-)[N])\\
     \nonumber &+ \kappa_y({\bf y}_+[1]-y_i) + \tilde \kappa_y( (\mathds{D}{\bf y}_+)[1]-\dot y_i)+ \xi_y ({\bf y}_-[N]) + \tilde \xi_y ((\mathds{D}{\bf y}_-)[N])
\end{align}\end{linenomath*}
\end{widetext}
As before, raising a discrete array to a power, as well as applying the exponential function acts element-wise in the above expression.

\end{document}